 \pdfoutput=1

\documentclass[10pt]{IEEEtran}
\usepackage{nopageno}

\usepackage{multirow}
\usepackage{booktabs}
\usepackage{tabularx}
\usepackage{ragged2e}
\usepackage{color}
\usepackage[utf8]{inputenc}
\usepackage{tabularx}
\usepackage{capt-of}
\usepackage[algo2e]{algorithm2e}
\usepackage{lipsum,multicol}
\usepackage{array,multirow}
\IEEEoverridecommandlockouts
\IEEEpubid{\makebox[\columnwidth]{978-1-5386-2880-5/17/\$31.00~\copyright~2017 IEEE \hfill} \hspace{\columnsep}\makebox[\columnwidth]{ }}

\PassOptionsToPackage{bookmarks={false}}{hyperref}
\usepackage{algorithm}
\usepackage{amsmath}
\usepackage{amsthm}
\usepackage{amssymb}
\usepackage{fancyvrb}
\usepackage{algorithmic}
\usepackage[T1]{fontenc}
\usepackage[scaled]{helvet}
\usepackage{lscape}
\usepackage[dvips,final]{graphicx}
\usepackage{subfigure}
\graphicspath{{./}{./figures/}}
\usepackage{enumerate}
\usepackage{epstopdf}
\usepackage{float}
\usepackage{url}
\usepackage{caption}
\usepackage{subfig}
\usepackage{booktabs}
\usepackage{cite}
\usepackage{array}

\newcommand\tabitem{\makebox[1em][r]{\textbullet~}}
\usepackage{tabularx}
\def\BibTeX{{\rm B\kern-.05em{\sc i\kern-.025em b}\kern-.08em T\kern-.1667em\lower.7ex\hbox{E}\kern-.125emX}}

\newcommand{\comment}[1]{ }

\IEEEoverridecommandlockouts

\begin{document}

\title{Traffic Behavior in Cloud Data Centers: A Survey\thanks{An IEEE-formatted version of this article is published in the  2020 International Wireless Communications and Mobile Computing (IWCMC) conference proceedings. Personal use of this material is permitted. Permission from IEEE must be obtained for all other uses, in any current or future media, including reprinting/republishing this material for advertising or promotional purposes, creating new collective works, for resale or redistribution to servers or lists, or reuse of any copyrighted component of this work in other works.}}


\author{Jarallah Alqahtani, Sultan Alanazi, Bechir Hamdaoui ~\\
\small Oregon State University, \small Corvallis,\\
\small \{alqahtaj,alanazsu,hamdaoui\}@eecs.oregonstate.edu~

}


\maketitle 
\begin{abstract}
Data centers (DCs) nowadays house tens of thousands of servers and switches, interconnected by high-speed communication links. With the rapid growth of cloud DCs, in both size and number, tremendous efforts have been undertaken to efficiently design the network and manage the traffic within these DCs. However, little effort has been made toward measuring, understanding and chattelizing how the network-level traffic of these DCs behave. In this paper, we aim to present a systematic taxonomy and survey of these DC studies. Specifically, our survey first decomposes DC network traffic behavior into two main stages, namely (1) data collection methodologies and (2) research findings, and then classifies and discusses the recent research studies in each stage. Finally, the survey highlights few research challenges related to DC network traffic that require further research investigation.

\end{abstract}
\thispagestyle{empty}


\section{\sc {Introduction}}
\label{sec:Introducation}
During the last decade, the intense usage of cloud-based services such as data storage, web services, and social media applications have significantly increased the data traffic within data center networks (DCNs). For instance, recent studies~\cite{cisco} project that global data traffic will reach 20.6 ZB by 2021, about a three-fold increase from 2016. Moreover, 73.3\% of this traffic is anticipated to stay within the data center (DC). As a result, significant research attention has recently been devoted to DCNs, ranging from designing new architectures and topologies to coming up with new performance metrics and developing methodologies for assessing and improving them. Despite these research efforts, there are limited studies that investigate and measure the traffic behavior of DCNs through real-world workloads. To preserve the Quality of Service (QoS) that DCNs offer their clients, a deeper measurement and analysis of the traffic behavior of the DCNs are crucial. For example, high packet drops and poor achievable throughputs caused by traffic congestions can lead to poor DCN performances (e.g., incur delay of search queries, increase failure rates of email services, and increase interruption rate of instant messaging).

These existing studies examine traces from real-world workloads in deployed networks. The studied data centers involve cloud, enterprise, and university domains, which differ not only in size and hosted application types, but also in traffic behaviorial aspects, including flow characteristics, presence and locality of burstiness, and traffic patterns. The traffic characteristics of these studies are highly correlated with DCN applications, which are responsible for introducing data into the network, with applications ranging from latency-sensitive ones such as Web search applications to throughput-sensitive ones such MapReduce applications.

\begin{figure*}[ht]
\centering
     \includegraphics[width=0.8\linewidth]{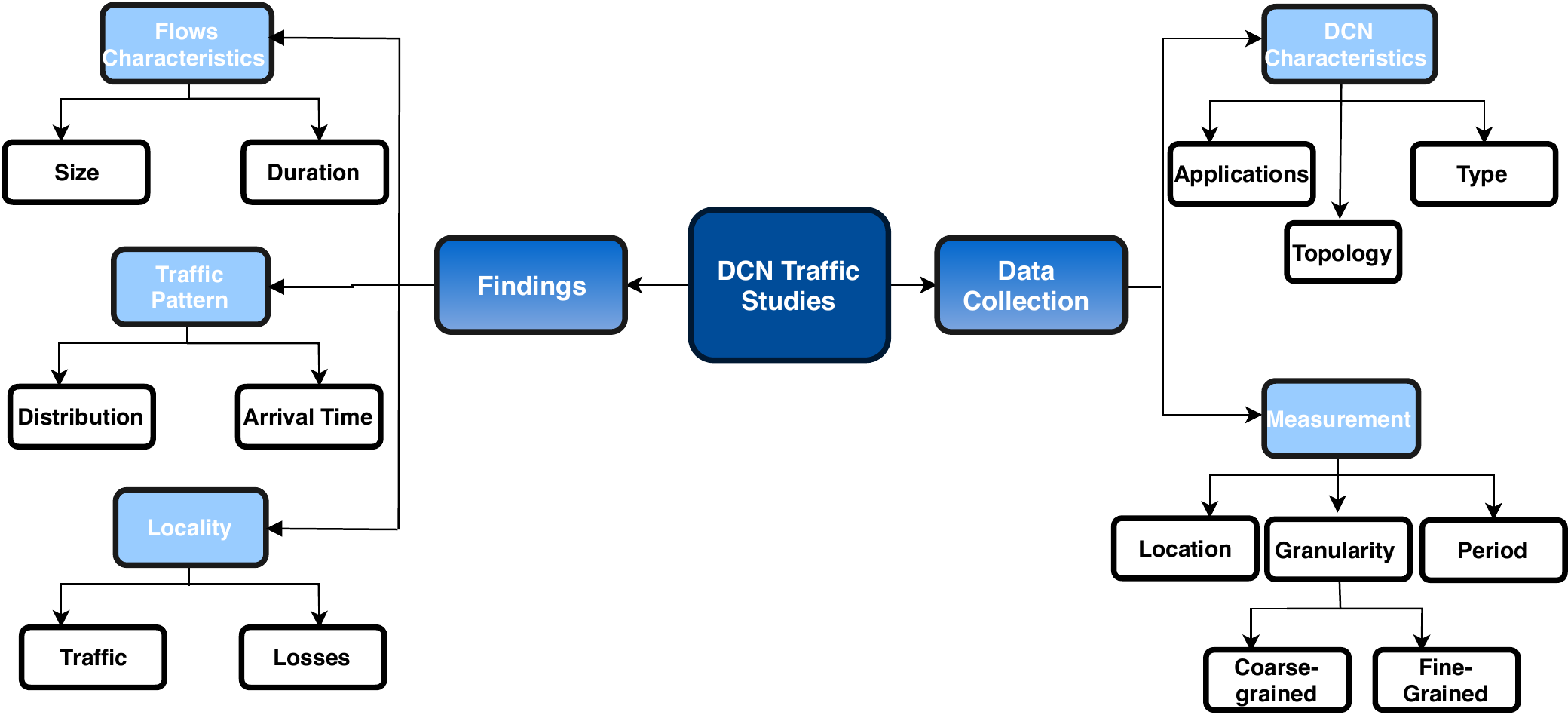}
      \caption{DCN Traffic Behavior Taxonomy.}
      \label{fig:1}
   \end{figure*}

In this paper, we present and discuss these studies in detail. To the best of our knowledge, this paper provides the first taxonomy and detailed comparison of DCN traffic behavior studies. As shown in Fig \ref{fig:1}, traffic characteristics of real DC workload studies conducted in literature can be decomposed into two main stages: 1) data collection, and 2) findings. Based on these two stages, we further break them down and classify them based on the features as presented Fig \ref{fig:1}.
The main contribution of this survey is the characterization of real workload DCN traffic studies by highlighting their similarities and differences. As a result, we believe this survey is a valuable resource for both researchers and practitioners seeking to understand DCN traffic behavior.

The remainder of this paper is organized as follows. Section \ref{sec:DataCollection} discusses and classifies existing studies based on the data collection method. In Section  \ref{sec:Findings}, we present the research findings of these studies, and classify the surveyed studies. We highlight some challenges and unsolved issues in DCN traffic behavior in Section \ref{sec:Discussion}, and conclude the paper in Section \ref{sec:Conclusion}.

\section{\sc {Data Collection}}
\label{sec:DataCollection}

In this section, we review and classify network behavior studies conducted on different data center networks based on their method of data collection. Furthermore, in Table \ref{DataCollection}, we compare in detail the methodology of collecting data for the surveyed studies in terms of deployed applications, DCN architecture, measurement tools and granularity.
Data collection plays a significant role in data center network traffic studies. A good study must present an accurate measurement and useful data that can be used in different type of data centers. The data used in current data center network studies vary based on DC characteristics and traffic measurement.

\subsection{DC Characteristics}
Three main parameters can affect DCN traffic behavior. These parameters, which we discuss next, are DC type, DCN topology, and DC applications.

\subsubsection{DC type}
\begin{figure}[ht]
     \includegraphics[width=\linewidth]{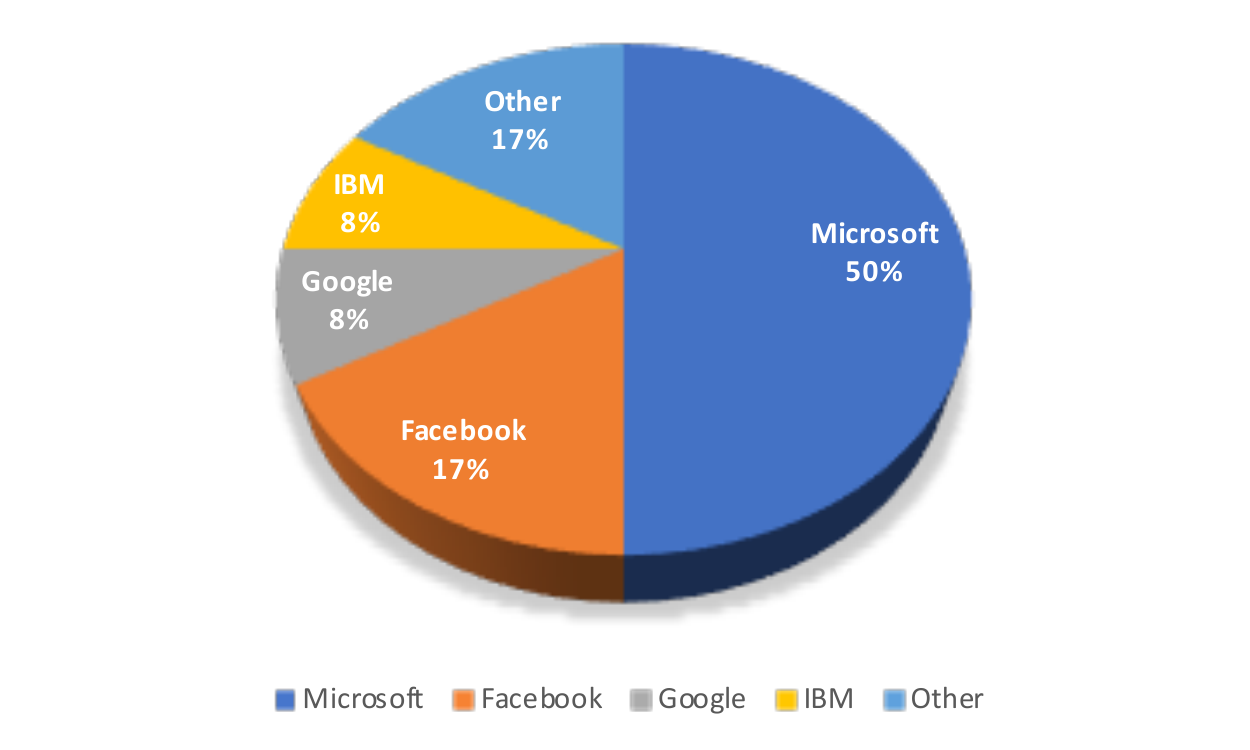}
      \caption{Percentage of studies conducted on different operational data centers .}
      \label{fig:fig2}
   \end{figure}

We classify studies under DC type into cloud and private data centers. Cloud DCs serve a variety of users and provide support for a broad range of internet services such as search, Email, video streaming, and real-time Messaging. Furthermore, cloud DCs have other systems hosted internally such as data mining, database, and storage to support the offered internet services. Some of these DCs are built with certain topology and oversubscription ratio to support specific applications, while other DCs are built for general purpose applications. We want to mention that the majority of traffic behavior studies are conducted on cloud DCs, namely Microsoft \cite{kandula2009nature}, Facebook \cite{roy2015inside}, and Google \cite{singh2015jupiter}, with, to the best of our knowledge, about 50$\%$ of the total studies being conducted on Microsoft DCs as shown in Fig~\ref{fig:fig2}.

On the other hand, private DCs serve a set of specific users and include universities and enterprises. University DCs serve students, faculty members, and administrative staff, and support a wide range of services such as student web portal, Email systems, administrative sites, and distributed file systems. Enterprise DCs, on the other hand, support a wide range of customized enterprise-specific applications along with the main applications like Web and Email services. There is only one study that focuses on private enterprise DC, IBM~\cite{meng2010improving}.
In addition, there are two DCN studies that focus on both cloud and private DCs, covering 19 DCs (\cite{benson2009understanding}) and 10 DCS (\cite{benson2010network}).

\subsubsection{DCN topology}
\begin{figure}[h]

     \includegraphics[width=\linewidth]{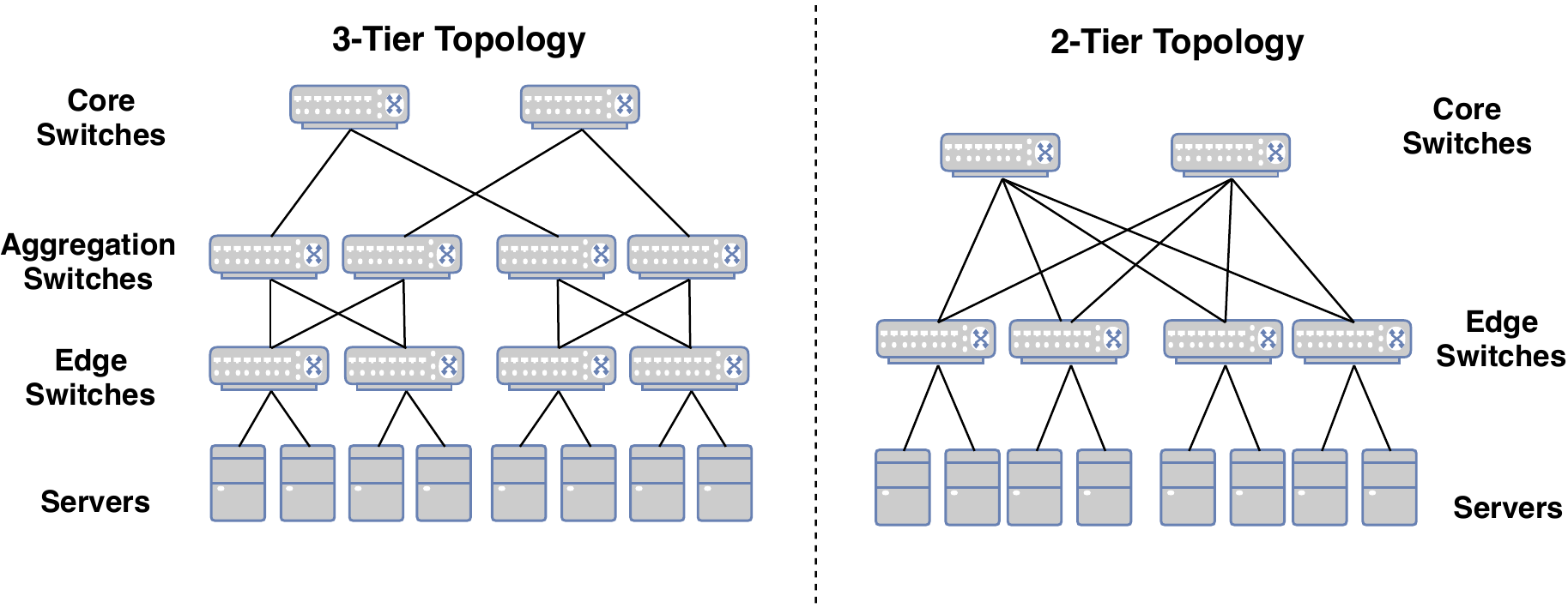}
      \caption{2-tier vs 3-tier Architecture.}
      \label{Arch}
\end{figure}

The Surveyed papers can be classified into two DC topology types: 3-tier and 2-tier. Fig \ref{Arch} shows the architecture of these two types. The 3 tiers of the 3-tier DC topology are the edge tier, which consists of the Top-of-Rack (ToR) switches that connect the servers to the DC's network fabric; the aggregation tier, which consists of devices that interconnect the ToR switches in the edge layer; and the core tier, which consists of devices that connect the DC to the WAN. Examples of studies conducted on 3-tier DCs include \cite{kandula2009nature,roy2015inside,zhang2017high,jalaparti2013speeding}.

In smaller DCs, the core tier and the aggregation tier are collapsed into one tier, resulting in a 2-tier DC topology. In current network behavior studies, 3-tier topology dominates cloud DC architectures while 2-tier is most common in private DCs \cite{benson2009understanding,benson2010network}.

\subsubsection{DC applications}
\begin{figure}[h]

     \includegraphics[width=\linewidth]{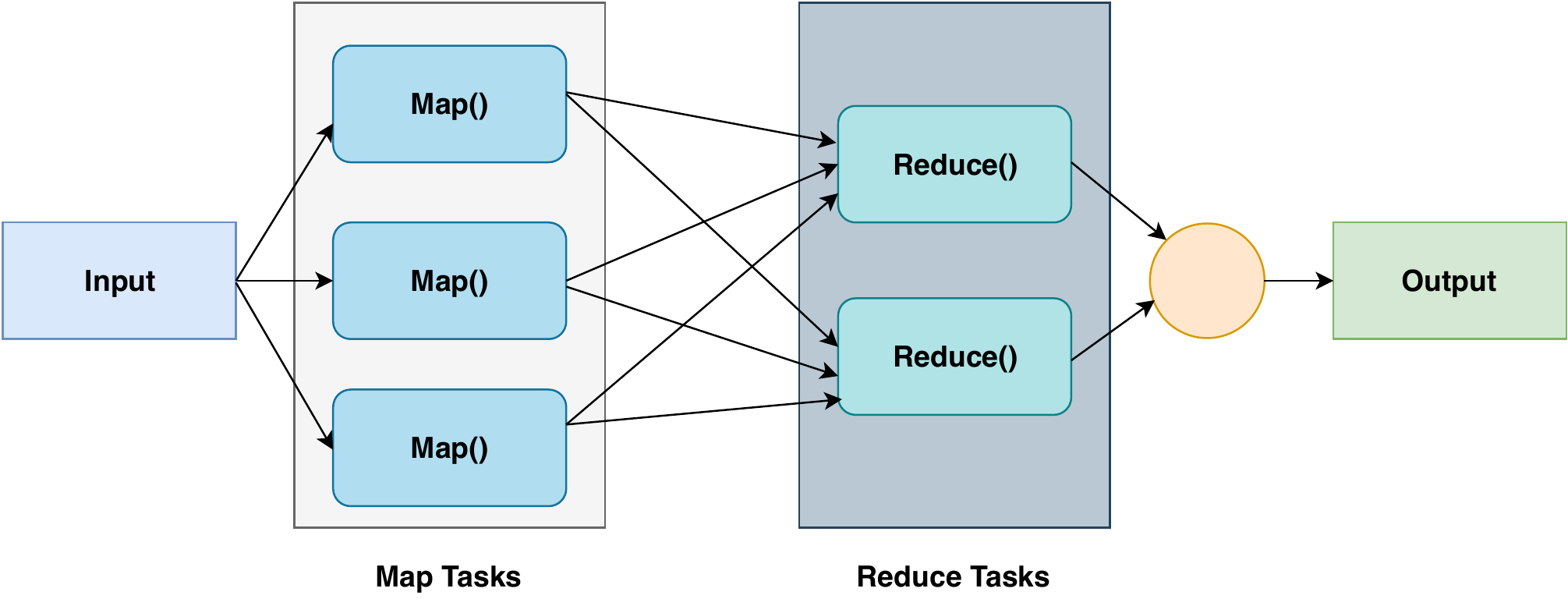}
      \caption{MapReduce.}
      \label{MapReduce}
\end{figure}

DCs support a variety of applications that result in distinct network traffic behaviors. We classify existing works based on DCN application into data mining, web services, and caches. MapReduce and Hadoop are the most popular applications of data mining.
MapReduce~\cite{dean2008mapreduce} is a programming model designed to handle data parallel applications by dividing the job into a set of independent tasks. It consists of two main functions: Map and Reduce. The Map function as shown in Fig \ref{MapReduce} generates a set of intermediate key/value pairs from large amount of row data. Then, the intermediate data get shuffled. The Reduce function as shown in Fig \ref{MapReduce} aggregates similar intermediate values and output the result. This model is widely used for system design. MapReduce is one key application that derives most of the traffic in the studied DC networks~\cite{benson2009understanding,benson2010network,kandula2009nature,greenberg2009vl2,kandula2009flyways,alizadeh2011data}.
Hadoop~\cite{shvachko2010hadoop} is another application used  for offline analysis and data mining. It is an efficient implementation of MapReduce, which provides both reliability and data transfer. Facebook researchers \cite{roy2015inside,zhang2017high}, for instance, collect traffic generated by Hadoop servers in their DCN traffic studies.

Web service applications, including web requests, Email, and video streaming, are also commonly used in DCs. The majority of traffic analysis papers capture and use traffic generated by web service applications \cite{benson2009understanding,benson2010network,roy2015inside,zhang2017high,delimitrou2012echo,alizadeh2011data,jalaparti2013speeding}. Cache serves as an in-memory cache of data used by the web servers. Some of these servers are leaders, which handle cache coherency, and some are followers, which serve most read requests. Such services constitute the main applications that generate traffic in Facebook DCs \cite{roy2015inside,zhang2017high}.

\subsection{Measurement}
Measurement can be decomposed into granularity and period. We classify studies based on their measurement granularity into coarse and fine-grained measurements. Coarse-grained is per-minute measurement and can be sampled using measurement tools such as  Simple Network Managment Protocol (SNMP)~\cite{SNMP} and Fbflow~\cite{roy2015inside}. The SNMP data  provide link-level statistics that can be used to study coarse-grained characteristics such as link utilization, packet drops, congestion, loss rates, etc. These statistics are collected by default in many DCs to detect major problems in the network. For example, the work in~\cite{benson2009understanding} presents an end-to-end traffic analysis by collecting tens of gigabytes of SNMP logs at 19 DCs over ten days to study coarse-grained characteristics of network traffic. Furthermore, authors of~\cite{benson2010network} collected SNMP link information in 10 DCs of three different organizations including cloud, university, and enterprise DCs for a period of ten days. SNMP traces are used in the studies to show the variation of link properties based on time and location in network-level.

Fbflow, on the other hand, is a monitoring tool that constantly samples packet headers across the whole network. It has two main components: Agents, which parse the headers of nflog samples extracting information including source and destination IP addresses, port numbers, and protocol type, and Taggers, which collect additional information such as the rack and cluster of sampled machines. Fbflow is deployed across the entire network of Facebook DCs~\cite{roy2015inside}, that connects hundreds of thousands of 10-Gbps nodes, to monitor data traffic among their major services.

Although these coarse-grained measurements are suitable for network monitoring and management, their sampling rate limits the study of some traffic properties including burstiness, and interarrival times that have an effect on traffic engineering approaches.

On the other hand, fine-grained is per-second measurements, such as port mirroring, packet sampling, and others. Port mirroring records packet-header traces over microsecond intervals. Facebook researchers~\cite{roy2015inside} turn on port mirroring on top of rack switches to capture the full, in-going and out-going traffic for selected servers.

Sampling packet directly using tcpdump is another way of providing fine-grained measurements such as packet inter-arrival times. Packet sampling can give a better understanding of traffic patterns. In~\cite{benson2009understanding} a part of their study is conducted on a small number of edge switches to obtain fine-grained traffic information. The data used in the study contains packet traces collected during a roughly two weeks period. Packet traces are collected from selected switches in private enterprise and university DCs that span 12 hours of multiple days~\cite{benson2010network}.  Packet traces allow authors to study different types of applications running in different DCs and the amount of traffic each application contributes to the network traffic. Furthermore, it facilitates examining the sending pattern at both packet and flow levels.


Researchers have proposed hardware modifications to provide more scalable and accurate measurements. However, these are not deployed widely enough to perform large-scale production measurements. Authors in~\cite{kandula2009nature} instrument servers to gather socket-level logs with minimal performance overhead. Over two months period, nearly a petabyte of measurements has been collected and analyzed from a 1500 server operational data center to report traffic patterns in terms of server communications and traffic characteristics in terms of flow statistics. This paper chooses a server-centric approach using Event Tracing for Windows (ETW)~\cite{ETW} to collect traffic events. The argument behind server-centric preference is that per-server monitoring incurs minimal overhead compared to network device monitoring.
Authors in~\cite{zhang2017high} focus mainly on fine-grained traffic behaviors on rack switches using a high-precision microburst measurement in data center networks. This paper modifies CPUs in modern switches to pool periodically local counters. Three sets of counters are the main focus in the paper which are Byte count, Packet size, and Peak buffer utilization. The measurement span 10 racks for each application type over one day period.

Obtaining finer grained data requires instrumentation of all switches and links in the cluster network; however, it is infeasible in today's DCs due to their large scale sizes.

\begin{table*}[ht]
\centering
\caption{Data Collection Comparison Table}
\label{DataCollection}
\begin{tabular}{|p{0.5cm}|p{2.0 cm}|p{1.0 cm}|p{1.5cm}|p{2.0cm}|p{1.5cm}|p{3.0cm}|p{2.5cm}|p{0.5cm}|}
\hline
\textbf{Paper} & \textbf{Type of DCN} &  \textbf{ \# of DCs }                            & \textbf{Topology}                  & \textbf{Applications} & \textbf{Duration} & \textbf{Measurement Granularity} &  \textbf{Measurement Tool} & \textbf{Year of Study} \\ \hline
      ~\cite{benson2009understanding} &  \tabitem Cloud  \newline   \tabitem Enterprise & 19 &  2 \& 3 tiers & \tabitem Web Services  \newline   \tabitem MapReduce         &   10 days                         &                   \tabitem Coarse-grained  \newline \tabitem  Limited Fine-grained           &                         \tabitem        SNMP   \newline   \tabitem Packet traces        &    2009           \\ \hline

      ~\cite{kandula2009nature} & \tabitem Cloud (Microsoft)                    & 1 & 3 tiers      &         \tabitem MapReduce     & 2 months        &        \tabitem Coarse-grained                      &   \tabitem SNMP   \newline   \tabitem Event Tracing for Windows (ETW)                 &   2009            \\ \hline

      ~\cite{benson2010network} &  \tabitem 5 Cloud  \newline   \tabitem 2  Enterprise \newline   \tabitem 3 University  & 10 &  2 \& 3 tiers  &  \tabitem Web Services  \newline   \tabitem MapReduce         &    10 days                         &     \tabitem Coarse-grained  \newline \tabitem  Limited Fine-grained        &              \tabitem SNMP  \newline  \tabitem Packet traces               &    2010           \\ \hline

     ~\cite{roy2015inside} &  \tabitem Cloud (Facebook)                    & 1 &   3 tiers &   \tabitem Web request \newline \tabitem cache \newline \tabitem Hadoop            &  One Day        &     \tabitem Coarse-grained  \newline \tabitem  Limited Fine-grained                &                      \tabitem        FbFlow \newline \tabitem
Port Mirroring  &               2015             \\ \hline

    ~\cite{zhang2017high}  & \tabitem Cloud (Facebook)                  & 1   &  3 tiers &      \tabitem Web request \newline \tabitem cache \newline \tabitem Hadoop       &    One day                 &         \tabitem   Limited Fine-grained                    &       \tabitem      CPUs in switches     &     2017          \\ \hline

     ~\cite{greenberg2009vl2} &  \tabitem Cloud (Microsoft)              & 1     &  3 tiers      &     \tabitem  MapReduce       &          &            \tabitem Coarse-grained             &       \tabitem       SNMP                         &  2009             \\ \hline

     ~\cite{alizadeh2011data} &  \tabitem Cloud (Microsoft)               & 1    &     3 tiers                      &        \tabitem  Web Services        &     One month     &           &                          &                                              2011      \\ \hline

      ~\cite{singh2015jupiter} &  \tabitem Cloud (Google)                  & 1     &  2 \& 3 tiers &              \tabitem Web Services  \newline   \tabitem MapReduce  &          &                          &                      &                    2015         \\ \hline

  ~\cite{delimitrou2012echo}    &  \tabitem Cloud (Microsoft)          & 1         &     3 tiers                      &     \tabitem Web Services         &     5 Months     & \tabitem  Limited Fine-grained                     &           &                                    2012             \\ \hline

     ~\cite{kandula2009flyways}  &  \tabitem Cloud (Microsoft)          & 1       &      3 tiers                     &     \tabitem MapReduce         &   Few months        &      \tabitem Coarse-grained                    &        \tabitem  Event Tracing for Windows (ETW)          &  2009              \\ \hline

    ~\cite{meng2010improving}  & \tabitem  Enterprise (IBM)           & 1        &                           &              &   10 days                      &         \tabitem Coarse-grained                     &                      &                    2010             \\ \hline

    ~\cite{jalaparti2013speeding}  & Cloud (Microsoft)        & 1         &  3 tiers      &      \tabitem Web Services  \newline   \tabitem MapReduce       &     Several hours    &                                  \tabitem  Limited Fine-grained       &                       \tabitem   Packet Traces    &                       2013  \\ \hline

\end{tabular}
\end{table*}

\section{\sc {Findings}}
\label{sec:Findings}

The existing studies of DCN traffic behavior can be categorized into three basic categories: flow characteristics, traffic pattern, and locality. 
For flow characteristics, studies show that both mice (small) and elephant (large) flows co-exist and are part of DC traffic. However, their existence varies based on DCN type and application. In addition to flow properties, traffic pattern is another key aspect of the DCN traffic behavior. Traffic patterns include flow arrival rates and distributions as well as traffic matrix. The traffic matrix represents how much traffic is exchanged between pairs of servers in the DC. Finally, locality of traffic, burstiness, and losses are getting much attention in the surveyed studies. Link utilization was measured to specify traffic amount in each layer. Packet drops also were quantitatively and spatially examined. We then explain these observations in detail and classify existing studies in Table~\ref{Findings} based on their findings.

\subsubsection{Flow Characteristics}

Flow characteristics including flow size and duration have been studied in some works, where DCN Traffic is frequently measured and characterized according to flows. A flow is a sequence of packets from a source to destination hosts. Flows can be large or small based on their sizes. Large flows are long-lived and throughput-sensitive while small flows are short-lived, latency sensitive, and highly bursty in nature. Many DC providers run a variety of applications that significantly influence the flow size. For example, flows that generated by web search applications are small in size (few KBs) while database update applications generate large flows ($>$ 1MB).

Most of DCN traffic behavior studies conclude that majority of flows are small in size ($\leq$ 10KB)~\cite{benson2010network,roy2015inside,greenberg2009vl2,alizadeh2011data}. These studies also exhibit lack of large flows and absence of super large flows. Moreover, most bytes are delivered by large flows~\cite{alizadeh2011data,greenberg2009vl2,benson2010network}. Besides the flow size, flow duration is also observed in some of these studies. According to~\cite{kandula2009nature}, most of flows duration are short. For example, about 80\% of flows last for less than 10 seconds. However, about .1\% of flows last longer than 200 seconds. In summary, there exist a mix of bandwidth-sensitive large flows and delay sensitive short flows in DC traffic.

\subsubsection{Traffic Patterns}

Understanding traffic patterns including flow arrival rates and distributions as well as traffic matrix are crucial for well designed DCNs.  Traffic matrix analysis allows DCN operators to allocate the necessary resources (bandwidth, VM placement, etc.) to meet the required quality of end-users experience. Unfortunately, it is difficult to build a robust traffic prediction system for DCN. First, because a detailed real traffic information needs to be collected in fine time granularity. Second, DCN currently deploys a wide variety of applications and these applications vary from simple web applications to more complex business workflow applications.

\begin{table}
\caption{Findings Comparison }
\label{Findings}
\begin{tabular}{|p{2.0cm}|p{4.2cm}|p{1.1cm}|}
\hline
\multicolumn{1}{|l|}{} & \multicolumn{1}{c|}{Findings} & Paper \\ \hline
\multirow{4}{*}  {\begin{tabular}[c]{@{}l@{}} \\ Fow\\  characteristics  \end{tabular}}& \tabitem Most of flows  in data centers are small in size (a few KB) \newline . & ~\cite{benson2010network,roy2015inside,greenberg2009vl2,alizadeh2011data}  \\ 
 & \tabitem Lack of large flows and absences for super large flows \newline & ~\cite{benson2010network,roy2015inside,greenberg2009vl2,kandula2009nature} \\
 & \tabitem Most bytes are delivered by large flows \newline  & ~\cite{alizadeh2011data,greenberg2009vl2,benson2010network}  \\
 & \tabitem Most of traffic bursts in the DCN are “microburst” \newline & ~\cite{zhang2017high}  \\ \hline
\multirow{9}{*} {Traffic Patterns}  & \tabitem Similarity in network activity patterns over time\newline & ~\cite{delimitrou2012echo}  \\
 & \tabitem ON/OFF behavior of packet arrivals \newline & ~\cite{benson2009understanding,benson2010network} \\
 & \tabitem Weak correlation between traffic rate and latency \newline &~\cite{meng2010improving}  \\
 & \tabitem Uneven distribution of traffic volumes from VMs \newline &  ~\cite{meng2010improving}\\ 
 & \tabitem Traffic changes over different time scales \newline & ~\cite{roy2015inside,delimitrou2012echo,benson2010network,kandula2009nature}\\
 & \tabitem Traffic pattern are unpredictable \newline & ~\cite{greenberg2009vl2}  \\
 & \tabitem Ratio of traffic volume between servers in data centers to traffic entering/leaving data centers 4:1  \newline & ~\cite{greenberg2009vl2} \\ 
 & \tabitem The inter-arrival time of  majority of microbursts are short \newline & ~\cite{zhang2017high}   \\
& \tabitem  Existence of Incast Problem \newline & ~\cite{alizadeh2011data,kandula2009nature}\\   \hline
\multirow{8}{*}  {\begin{tabular}[c]{@{}l@{}} \\ \\ \\  Locality of Traffic\\  and Losses \end{tabular}}  & \tabitem Packet,drops are highest at the edge \newline & ~\cite{benson2009understanding,benson2010network,zhang2017high}  \\ 
 &  \tabitem Links utilization in the core level are the highest \newline & ~\cite{benson2009understanding,benson2010network,roy2015inside} \\ 
 &  \tabitem Majority of cloud data centers traffic stays within the rack \newline & ~\cite{benson2010network,kandula2009nature,delimitrou2012echo}  \\
 &  \tabitem Small fraction of discards is due to oversubscription of ToR uplinks towards the fabric \newline & ~\cite{singh2015jupiter}  \\ 
 &  \tabitem ToR switches layer are more bursty \newline & ~\cite{zhang2017high}   \\
 &  \tabitem Only a few ToRs are hot and most of their traffic goes to a few other ToRs \newline & ~\cite{kandula2009flyways} \\ 
 &  \tabitem Traffic is not localized to any particular layer and it depends upon the service \newline &   ~\cite{roy2015inside} \\ 
 &  \tabitem Traffic locality is stable across time period \newline & ~\cite{roy2015inside,delimitrou2012echo} \\ \hline
\end{tabular}
\end{table}

However, several studies of real workload exploited some basic traffic characteristics to describe traffic pattern of cloud DCs. For example, ~\cite{benson2010network,benson2009understanding}  shows that packet arrivals evidence an ON/OFF traffic behavior after packet traces of private enterprise and university DCs. Furthermore, this pattern fits best with log-normal distributions. Another study of Microsoft DCN traffic~\cite{greenberg2009vl2} concludes that total traffic matrix of cloud DCN is unpredictable and fluctuating. In spite of that, a study of Facebook DCN traffic pattern~\cite{roy2015inside} concludes that traffic strongly exhibits continuous arrivals and do not evidence on/off pattern.
In addition, it is observed that DCN traffic can be very bursty, and most of this bursty traffic is “microburst”. For example, about 90 $\%$ of  bursts  last  less than 200[$\mu s$]. Moreover, the inter-arrival time of these majority $\mu$ bursts are short. For example,  inter-bursts last less than 100$\mu$ for about 40$\%$ of Cache and Web services.

\subsubsection{Locality of Traffic and Losses}
Another key aspect of traffic properties of DCN is traffic and loss localization. Determining the spatial locality of traffic and packet losses also advocates DCN performance efficiency.  Although the lack of predictable traffic matrix can provide no locality of traffic, DCN traffic studies evidence some locality trends of traffic and losses. First, surprisingly, there is a weak correlation between link utilization and localization of packet drops. For example,~\cite{benson2009understanding,benson2010network,roy2015inside} observe that link utilization in the core level is higher than other levels. However, packet drops are highest at the edge~\cite{benson2009understanding,benson2010network,zhang2017high}. The reason behind this loss is that traffic burstiness is higher at the rack level. In~\cite{benson2010network,kandula2009nature,delimitrou2012echo}, authors noted that most traffic in the cloud is restricted to within a rack. In contrast, a significant fraction of traffic of the enterprise and universities DCs leaves the rack~\cite{benson2009understanding,benson2010network}. Study on Facebook DCN~\cite{roy2015inside} found that traffic is not localized to any particular layer and depends on the offered service. For example, in Hadoop server's majority of traffic is intra-rack while traffic in Web servers are largely intra-cluster. In addition to heavy racks traffic observation, most of their traffic goes to a few other ToRs~\cite{kandula2009flyways}. In other words, there are few ToRs that are likely prone to be bottleneck. To recap, locality of traffic and losses are determined by applications that run within the DCN and how these applications are deployed on the servers.


To sum up, DCN traffic characteristics, including flow size and distribution, locality and episode of burstness, are highly dependant on applications. In addition, most flows in DCs are small in size (a few KBs). However, large flows whose length exceeds a few hundreds MB rarely exist. Moreover, plenty of these studies conclude that the traffic patterns are highly rack local. The reason behind this behavior is that DCN operators place the applications in such a way that the inter-rack traffic is reduced. Finally,  although  traffic  patterns originated from a rack exhibit an ON/OFF pattern traffic and fit best with log-normal distributions, the entire traffic matrix is random and hardly predictable. 


%
%
%
\section{\sc {Discussion}}
\label{sec:Discussion}
\subsection{A Multi-Tenant Data Center (MTDC) traffic}
	
An infrastructure as a service (IaaS) cloud providers, such as Amazon AWS~\cite{AWS}, Microsoft Azure~\cite{AZURE}, and Google Compute Engine~\cite{Google}, offers computing and storage resources to tenants (customers) by hosting their computing instances on the cloud provider's physical machines. In a traditional single-tenant cloud DC, providers offer a dedicated cloud service to their tenants, where resources are not shared with other tenants. During the last decade, many organizations choose to move and obtain their operations, storage and computation to multi-tenant (MT) DCs. However, we observed that only one paper (IBM) briefly studied network traffic behavior on MTDC, but the rest considers single-tenant DCs. We believe MTDCs exhibit traffic behaviors that are different from those of single-tenant DCs.

\subsection{Implications for Load Balancing}

Majority of the surveyed studies use the Equal-Cost Multiple Path (ECMP)~\cite{hopps2000analysis} approach to load balance flows among ToRs. However, only a couple studies briefly mentions the implication of load balancing mechanisms. We believe that traffic congestion and loses observed in these studies are mainly due to their deployed load balancing mechanisms.  It is well known that ECMP performs poorly in DC networks due to hash collisions. ECMP can cause congestions when two long-running flows are assigned to the same path. Also, ECMP doesn't adapt well to asymmetry in the network topology. As a result,  a rich body of work such as Hedera~\cite{hedera}, Conga~\cite{conga}, FastPass~\cite{fastpass}, Clove~\cite{clove}, Presto~\cite{presto}, Luopan~\cite{luopan},  have been proposed in literature  to address the problem of load balancing in DCN. However, none of these schemes are used in today's DCs. Future DCN traffic studies should investigate the implication of different load balancing schemes on the traffic behavior of DCs.

\subsection{High Resolution Measurement}
Real-time applications such as Web services are delay sensitive and strictly require QoS to be maintained. As a result, measuring and monitoring of real-time traffic at high-speed DCNs is substantial. Moreover, it is important for evaluating new protocols and architectures.  Current studies instrument a small fraction of network devices for short time scale (few hours) to obtain fine-grained measurement, which does not reflect the entire traffic behavior of high speed DCs.

\subsection{Lack of Traces}
Traffic traces represent a very rich and useful source of information not only for understanding the traffic characteristics, but also in evaluating newly proposed solution approaches for DCs. Furthermore, due to the lack of these real traces, the majority of the proposed solutions for efficient DCN including topology design and traffic management are evaluated based on synthetic data, which is not a perfect representative of large-scale production workloads. 
There is a clear need for measuring and collecting real traces of DC traffic and especially for making them publicly available for DC researchers.

\section{\sc {Conclusion}}
\label{sec:Conclusion}
Different studies for measuring and understanding operational data center network behavior have been conducted during last decade. We review and classify these works focusing on the way they collect the traffic as well as the key observations made from these studies.  We have learned that it is difficult to conduct entire fin-grained measurements for large scale data centers that produce terabytes of traffic per second. We also learned that DCN traffic typically correlated with applications and exhibit irregular and volatile patterns. 

\bibliographystyle{IEEEtran}
\bibliography{References}

\begin{thebibliography}{10}
\providecommand{\url}[1]{#1}
\csname url@samestyle\endcsname
\providecommand{\newblock}{\relax}
\providecommand{\bibinfo}[2]{#2}
\providecommand{\BIBentrySTDinterwordspacing}{\spaceskip=0pt\relax}
\providecommand{\BIBentryALTinterwordstretchfactor}{4}
\providecommand{\BIBentryALTinterwordspacing}{\spaceskip=\fontdimen2\font plus
\BIBentryALTinterwordstretchfactor\fontdimen3\font minus
  \fontdimen4\font\relax}
\providecommand{\BIBforeignlanguage}[2]{{%
\expandafter\ifx\csname l@#1\endcsname\relax
\typeout{** WARNING: IEEEtran.bst: No hyphenation pattern has been}%
\typeout{** loaded for the language `#1'. Using the pattern for}%
\typeout{** the default language instead.}%
\else
\language=\csname l@#1\endcsname
\fi
#2}}
\providecommand{\BIBdecl}{\relax}
\BIBdecl

\bibitem{cisco}
C.~V. networking Index, ``Forecast and methodology, 2016-2021, white paper,''
  \emph{San Jose, CA, USA}, vol.~1, 2016.

\bibitem{kandula2009nature}
S.~Kandula, S.~Sengupta, A.~Greenberg, P.~Patel, and R.~Chaiken, ``The nature
  of data center traffic: measurements \& analysis,'' in \emph{Proceedings of
  the 9th ACM SIGCOMM conference on Internet measurement}.\hskip 1em plus 0.5em
  minus 0.4em\relax ACM, 2009, pp. 202--208.

\bibitem{roy2015inside}
A.~Roy, H.~Zeng, J.~Bagga, G.~Porter, and A.~C. Snoeren, ``Inside the social
  network's (datacenter) network,'' in \emph{ACM SIGCOMM Computer Communication
  Review}, vol.~45, no.~4.\hskip 1em plus 0.5em minus 0.4em\relax ACM, 2015,
  pp. 123--137.

\bibitem{singh2015jupiter}
A.~Singh, J.~Ong, A.~Agarwal, G.~Anderson, A.~Armistead, R.~Bannon, S.~Boving,
  G.~Desai, B.~Felderman, P.~Germano \emph{et~al.}, ``Jupiter rising: A decade
  of clos topologies and centralized control in google's datacenter network,''
  in \emph{ACM SIGCOMM Computer Communication Review}, vol.~45, no.~4.\hskip
  1em plus 0.5em minus 0.4em\relax ACM, 2015, pp. 183--197.

\bibitem{meng2010improving}
X.~Meng, V.~Pappas, and L.~Zhang, ``Improving the scalability of data center
  networks with traffic-aware virtual machine placement,'' in \emph{INFOCOM,
  2010 Proceedings IEEE}.\hskip 1em plus 0.5em minus 0.4em\relax IEEE, 2010,
  pp. 1--9.

\bibitem{benson2009understanding}
T.~Benson, A.~Anand, A.~Akella, and M.~Zhang, ``Understanding data center
  traffic characteristics,'' in \emph{Proceedings of the 1st ACM workshop on
  Research on enterprise networking}.\hskip 1em plus 0.5em minus 0.4em\relax
  ACM, 2009, pp. 65--72.

\bibitem{benson2010network}
T.~Benson, A.~Akella, and D.~A. Maltz, ``Network traffic characteristics of
  data centers in the wild,'' in \emph{Proceedings of the 10th ACM SIGCOMM
  conference on Internet measurement}.\hskip 1em plus 0.5em minus 0.4em\relax
  ACM, 2010, pp. 267--280.

\bibitem{zhang2017high}
Q.~Zhang, V.~Liu, H.~Zeng, and A.~Krishnamurthy, ``High-resolution measurement
  of data center microbursts,'' in \emph{Proceedings of the 2017 Internet
  Measurement Conference}.\hskip 1em plus 0.5em minus 0.4em\relax ACM, 2017,
  pp. 78--85.

\bibitem{jalaparti2013speeding}
V.~Jalaparti, P.~Bodik, S.~Kandula, I.~Menache, M.~Rybalkin, and C.~Yan,
  ``Speeding up distributed request-response workflows,'' in \emph{ACM SIGCOMM
  Computer Communication Review}, vol.~43, no.~4.\hskip 1em plus 0.5em minus
  0.4em\relax ACM, 2013, pp. 219--230.

\bibitem{dean2008mapreduce}
J.~Dean and S.~Ghemawat, ``Mapreduce: simplified data processing on large
  clusters,'' \emph{Communications of the ACM}, vol.~51, no.~1, pp. 107--113,
  2008.

\bibitem{greenberg2009vl2}
A.~Greenberg, J.~R. Hamilton, N.~Jain, S.~Kandula, C.~Kim, P.~Lahiri, D.~A.
  Maltz, P.~Patel, and S.~Sengupta, ``Vl2: a scalable and flexible data center
  network,'' in \emph{ACM SIGCOMM computer communication review}, vol.~39,
  no.~4.\hskip 1em plus 0.5em minus 0.4em\relax ACM, 2009, pp. 51--62.

\bibitem{kandula2009flyways}
S.~Kandula, J.~Padhye, and P.~Bahl, ``Flyways to de-congest data center
  networks,'' 2009.

\bibitem{alizadeh2011data}
M.~Alizadeh, A.~Greenberg, D.~A. Maltz, J.~Padhye, P.~Patel, B.~Prabhakar,
  S.~Sengupta, and M.~Sridharan, ``Data center tcp (dctcp),'' \emph{ACM SIGCOMM
  computer communication review}, vol.~41, no.~4, pp. 63--74, 2011.

\bibitem{shvachko2010hadoop}
K.~Shvachko, H.~Kuang, S.~Radia, and R.~Chansler, ``The hadoop distributed file
  system,'' in \emph{Mass storage systems and technologies (MSST), 2010 IEEE
  26th symposium on}.\hskip 1em plus 0.5em minus 0.4em\relax Ieee, 2010, pp.
  1--10.

\bibitem{delimitrou2012echo}
C.~Delimitrou, S.~Sankar, A.~Kansal, and C.~Kozyrakis, ``Echo: Recreating
  network traffic maps for datacenters with tens of thousands of servers,'' in
  \emph{Workload Characterization (IISWC), 2012 IEEE International Symposium
  on}.\hskip 1em plus 0.5em minus 0.4em\relax IEEE, 2012, pp. 14--24.

\bibitem{SNMP}
J.~Case, R.~Mundy, D.~Partain, and B.~Stewart, ``Introduction and applicability
  statements for internet-standard management framework,'' Tech. Rep., 2002.

\bibitem{ETW}
``Event tracing for windows (etw),''
  \url{https://docs.microsoft.com/en-us/windows/desktop/etw/event-tracing-portal},
  note = {Accessed: 2020-02-20}.

\bibitem{AWS}
``Amazon aws kernel description,'' \url{https://aws.amazon.com/}, accessed:
  2020-02-20.

\bibitem{AZURE}
``Microsoft azure,'' \url{https://azure.microsoft.com/}, accessed: 2020-02-20.

\bibitem{Google}
``Google compute engine,'' \url{https://cloud.google.com/compute/}, accessed:
  2020-02-20.

\bibitem{hopps2000analysis}
C.~Hopps, ``Analysis of an equal-cost multi-path algorithm,'' Tech. Rep., 2000.

\bibitem{hedera}
M.~Al-Fares, S.~Radhakrishnan, B.~Raghavan, N.~Huang, and A.~Vahdat, ``Hedera:
  dynamic flow scheduling for data center networks.'' in \emph{Nsdi}, vol.~10,
  no.~8, 2010, pp. 89--92.

\bibitem{conga}
M.~Alizadeh, T.~Edsall, S.~Dharmapurikar, R.~Vaidyanathan, K.~Chu,
  A.~Fingerhut, F.~Matus, R.~Pan, N.~Yadav, G.~Varghese \emph{et~al.}, ``Conga:
  Distributed congestion-aware load balancing for datacenters,'' in \emph{ACM
  SIGCOMM Computer Communication Review}, vol.~44, no.~4.\hskip 1em plus 0.5em
  minus 0.4em\relax ACM, 2014, pp. 503--514.

\bibitem{fastpass}
J.~Perry, A.~Ousterhout, H.~Balakrishnan, D.~Shah, and H.~Fugal, ``Fastpass: A
  centralized zero-queue datacenter network,'' \emph{ACM SIGCOMM Computer
  Communication Review}, vol.~44, no.~4, pp. 307--318, 2015.

\bibitem{clove}
N.~Katta, M.~Hira, A.~Ghag, C.~Kim, I.~Keslassy, and J.~Rexford, ``Clove: How i
  learned to stop worrying about the core and love the edge,'' in
  \emph{Proceedings of the 15th ACM Workshop on Hot Topics in Networks}.\hskip
  1em plus 0.5em minus 0.4em\relax ACM, 2016, pp. 155--161.

\bibitem{presto}
K.~He, E.~Rozner, K.~Agarwal, W.~Felter, J.~Carter, and A.~Akella, ``Presto:
  Edge-based load balancing for fast datacenter networks,'' in \emph{ACM
  SIGCOMM Computer Communication Review}, vol.~45, no.~4.\hskip 1em plus 0.5em
  minus 0.4em\relax ACM, 2015, pp. 465--478.

\bibitem{luopan}
P.~Wang, G.~Trimponias, H.~Xu, and Y.~Geng, ``Luopan: Sampling-based load
  balancing in data center networks,'' \emph{IEEE Transactions on Parallel and
  Distributed Systems}, vol.~30, no.~1, pp. 133--145, 2019.

\end{thebibliography}
\end{document}